\documentclass[conference]{IEEEtran}
\IEEEoverridecommandlockouts
\usepackage{cite}
\usepackage{amsmath,amssymb,amsfonts}
\usepackage{subcaption}
\usepackage{algorithmic}
\usepackage{graphicx}
\usepackage{textcomp}
\usepackage{xcolor}
\usepackage{hyperref}
\usepackage[capitalize]{cleveref}
\usepackage{tikz}
\usetikzlibrary{arrows.meta, positioning}
\def\BibTeX{{\rm B\kern-.05em{\sc i\kern-.025em b}\kern-.08em
    T\kern-.1667em\lower.7ex\hbox{E}\kern-.125emX}}
\begin{document}

\title{Symbolic Timing Analysis of Digital Circuits Using Analytic Delay Functions
\\
}
\author{
\IEEEauthorblockN{Era Thaqi\IEEEauthorrefmark{1}, Dennis Eigner\IEEEauthorrefmark{2}, Arman Ferdowsi\IEEEauthorrefmark{3}\IEEEauthorrefmark{2}, Ulrich Schmid\IEEEauthorrefmark{2}}
\IEEEauthorblockA{\IEEEauthorrefmark{1}Electrical and Computer Engineering, Princeton University, Princeton, NJ, USA \\
era.thaqi@princeton.edu}
\IEEEauthorblockA{\IEEEauthorrefmark{2}Embedded Computing Systems Group, TU Wien, Vienna, Austria \\
e11808235@student.tuwien.ac.at, arman.ferdowsi@tuwien.ac.at, s@ecs.tuwien.ac.at}
\IEEEauthorblockA{\IEEEauthorrefmark{3}Faculty of Computer Science, University of Vienna, Vienna, Austria \\
arman.ferdowsi@univie.ac.at}
}

\maketitle

\begin{abstract}
We propose a novel approach to symbolic timing analysis for digital integrated circuits based on recently developed analytic delay formulas for 2-input NOR, NAND, and Muller-C gates by Ferdowsi et~al. (NAHS 2025). Given a fixed order of the transitions of all input and internal signals of a circuit, our framework computes closed-form analytic delay expressions for all the internal signal transition times that depend on (i) the symbolic transition times of the relevant input signals and (ii) the model parameters of the relevant gates. The resulting formulas facilitate per-transition timing analysis without any simulation, by instantiating the symbolic input transition times and the gate parameters. More importantly, however, they also enable an \emph{analytic} study of the dependencies of certain timing properties on input signals and gate parameters. For instance, differentiating a symbolic delay expression with respect to a gate parameter or input transition time enables sensitivity analysis.
As a proof of concept, we implement our approach using the computer algebra system SageMath and apply it to the NOR-gate version of the c17 slack benchmark circuit. 

\end{abstract}

\begin{IEEEkeywords}
symbolic timing analysis, digital circuit verification, analytic delay models, computer algebra tool
\end{IEEEkeywords}

\section{Introduction}

Recent analytic delay models for 2-input logic gates, notably NOR, NAND, and Muller C gates, have yielded delay functions $\delta(T,\Delta)$ that capture both (i) drafting effects, where a gate's output delay depends on the time since its previous transition, and (ii) multi-input switching (MIS) effects arising from temporally proximate changes of different inputs. These models, introduced by Ferdowsi et al.~\cite{ferdowsi2025}, offer closed-form expressions derived from a thresholded hybrid ODE representation of CMOS transistor behaviors, and have been experimentally shown to provide excellent accuracy in comparison to gold standard SPICE-generated traces, which are extensively computationally expensive.

In this work, we present a symbolic framework that composes $\delta(T,\Delta)$ expressions of the gates in a circuit for a given fixed sequence of input transitions: Based on the transition order of all signals in the circuit specified by the user, our algorithm computes \emph{output transition times} in symbolic form, using formulas that involve symbolic input transition times and gate parameters. After instantiating these symbolic values, one can check whether the assumed transition ordering is \emph{consistent} with the actual delays. Our framework can thus be used as a validation mechanism for candidate timing paths, facilitating the exploration of delay dependencies under varying input timing and parameter regimes. For example, one can symbolically determine input timing constraints that ensure a particular ordering of all the signals in the circuit. Similarly, the tool can also be used to symbolically evaluate how physical gate parameters influence the transition ordering. The closed-form constraints can be directly embedded in SAT/SMT-based timing proofs, thereby linking transistor-accurate delay modeling with the formal verification workflows.

\begin{figure}[htbp]
    \centering
    \includegraphics[width=0.75\columnwidth]{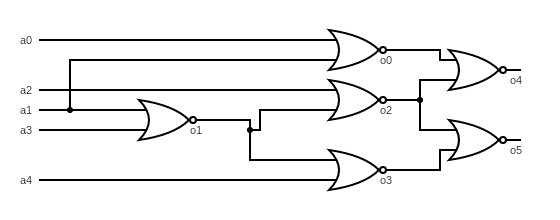}
    \caption{ \small{C17 with NOR gates.}}
    \label{fig:c17}
\end{figure}

\section{Prototype and Case Study}

So far, the delay models developed in \cite{ferdowsi2025} have been implemented and used only for dynamic digital
timing simulation~\cite{ferdowsi2025timing}. The gate delay functions employed there are based on a hybrid model, which internally employs
expressions for the analog gate output voltage $V_{\text{int}}$. However, recent work has shown that it is possible 
to completely eliminate $V_{\text{int}}$ and encode its effect entirely through the ordering and timing of (i) some input transitions (captured by some parameter $\Delta$) and (ii) their relation to the previous output transition (captured by some parameter $T$). Since MIS effects depend on the particular ordering and direction of the input transitions, different cases (a),\dots,(g) for input transitions must be distinguished. The result are fully digital gate delay expressions $\delta(T, \Delta)$ that capture drafting and MIS effects without any reference to analog voltage trajectories, for every two-step transition sequence (x,y). Our framework exploits these delay formulas: by tracking both the initiating and competing transitions at each gate input, it facilitates symbolic propagation of delays through arbitrary circuits using only discrete event sequences.

We implemented our symbolic timing analysis framework using SageMath as the embedded Computer Algebra Tool. It accepts discrete input transition sequences with symbolic occurrence times and a circuit description. For each gate in the circuit, it dynamically determines the relevant transition cases and applies the appropriate delay formulas for computing the symbolic output transition times. In the case of gate inputs driven by the output of another gate, this needs to be done in an appropriately nested fashion. Note that this works even for circuits involving feedback loops, where symbolic output transition times involve previous symbolic output transition times, i.e., are recursive.

Using the case of \cref{fig:c17} as an example, starting from the symbolic input transitions, the framework computes symbolic expressions for the outputs o0 through o5, respecting the gate-level timing dependencies. Each output is computed as a closed-form analytic function of the prior transitions and gate parameters. 

\begin{figure}[htbp]
\centering
\begin{tikzpicture}[scale=0.8, every node/.style={scale=0.95}]

  \node at (0, 4) {\small \textcolor{blue}{(0,0)}};
  \draw[->, black, thick] (0.3,4) -- (0.9,4) node[midway, above] {\textbf{a}};
  \node at (1.2, 4) {\small \textcolor{blue}{(1,0)}};
  \draw[->, black, thick] (1.5,4) -- (2.1,4) node[midway, above] {\textbf{c}};
  \node at (2.4, 4) {\small \textcolor{blue}{(1,1)}};
  \draw[->, black, thick] (2.7,4) -- (3.3,4) node[midway, above] {\textbf{e}};
  \node at (3.6, 4) {\small \textcolor{blue}{(0,1)}};
  \draw[->, black, thick] (3.9,4) -- (4.5,4) node[midway, above] {\textbf{g}};
  \node at (4.8, 4) {\small \textcolor{blue}{(0,0)}};

  \node[left] at (-0.2, 3) {\textbf{a1}};
  \draw[green!60!black, very thick] (0,3) -- (1,3) -- (1,3.3) -- (3,3.3) -- (3,3) -- (6,3);

  \node[left] at (-0.2, 2) {\textbf{a3}};
  \draw[green!60!black, very thick] (0,2) -- (2.5,2) -- (2.5,2.3) -- (4,2.3) -- (4,2) -- (6,2);

  \node[left] at (-0.2, 1) {\textbf{o1}};
  \draw[red, very thick] (0,1) -- (1.7,1) -- (1.7,0.6) -- (5,0.6) -- (5,1) -- (6,1);

  \draw[<->, thick, blue] (1,0.5) -- (1.7,0.5);
  \node[blue] at (1.35,0.7) {\scriptsize $\delta(a)$};

  \draw[<->, thick, orange] (1.7,1) -- (2.5,1);
  \node[orange] at (2.1,1.2) {\scriptsize $T = t_1 - t_{o1}$};

  \node at (1, 2.7) {$t_0$};
  \node at (2.5, 1.7) {$t_1$};
  \node at (3, 2.7) {$t_2$};
  \node at (4, 1.7) {$t_3$};

  \node[black, font=\bfseries] at (2.75,0.4) {\scriptsize $t_{o1} = \delta(a) + t_0$};

  \draw[thick] (0,-0.2) -- (6.5,-0.2);

\end{tikzpicture}
    \caption{ \small{Timing diagram showing the first formula $t_{o1}$ symbolically.} }
    \label{fig:symbolic_timing}
\end{figure}
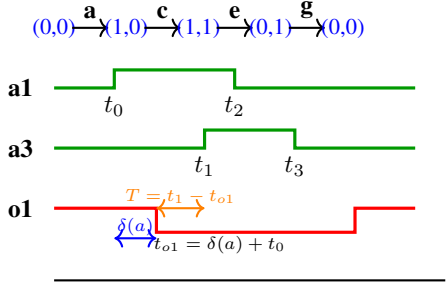

To illustrate the inner workings of the approach, \cref{fig:symbolic_timing} shows a transition sequence (a),(c),(e),(g), with the corresponding two-step transition sequence (a,c), (c,e), (e,g), where e.g., (a,c) represents the transitions case (a) followed by case (c), and the resulting first internal output $o1$. Out tool combines the gate delay formulas of the respective cases to compute the closed-form occurrence time of $o1$.

The formula below illustrates the delay for case (a,c) \cite{ferdowsi2025timing}:
\begin{align}
\text{Case (a,c):} \quad
\delta^{\downarrow}(T) &= \frac{-C_2 R_{nB} (T + \delta_{\min})}{C_1(R_{nA} + R_{nB})} + \delta_{\min}
\end{align}

Subsequent output nodes (e.g., $o_1, o_2, o_4$) are handled similarly, using the computed transitions of their fan-in gates as symbolic inputs. For example, output $o_1$ depends on $t_0$, the initial external signal, and recursively built expressions for later transitions.

The framework proceeds gate by gate, transition by transition, without resorting to numeric simulation or evaluation. Our fully symbolic pipeline thus preserves the precise algebraic structure of timing dependencies in a circuit, making it possible to reason about parameter sensitivities, propagation patterns, and timing margin violations under variation.

\section{Applications}

Our framework enables symbolic exploration of gate-level timing behaviors in a circuit. Because transition times are computed as symbolic expressions, we can perform analytic parameter sensitivity analysis, e.g., evaluating the conditions for some gate parameters or input transition time differences that maximize or minimize certain output delays. For example, it is known that the self-timed ring oscillator described in \cite{winstanley2002} 
can exhibit two different operation modes, namely, equally spaced output transitions ($t_{n+1} - t_n = t_n - t_{n-1}$) 
or burst mode ($t_{n+1} - t_n \neq t_n - t_{n-1}$), which can even alternate in a chaotic fashion. The exact conditions on the gate parameters that causes these modes and the mode switches are unknown, though. The recursive formulas for $t_n$ provided by our tool should allow us to determine those conditions.

We also envision a possible application of our framework for path validation in automated verification pipelines. Given a proposed timing permutation, the tool can symbolically compute all delays and verify whether the assumed ordering is consistent with the resulting arrival times. This could provide an efficient alternative to brute-force enumeration or exhaustive SPICE simulations in validating timing paths.

\section{Future Work}

A separate line of future work targets the biggest current limitation of our approach: the assumption of the a priori given transition order of all signals in a circuit. Our aim is to eliminate this requirement by coupling our symbolic timing analysis tool with a symbolic execution framework, which determines the relevant transition orderings. For each candidate order, our tool would compute the corresponding timing expressions. This can be done optimistically, as infeasible orders can be pruned when their symbolic constraints become unsatisfiable; the remaining orders would reveal the truly critical paths. 

The symbolic execution framework we intend to use constructs a tree of possible transition orderings and uses heuristics to select the most promising branches without requiring backtracking. Rather than exhaustively exploring the entire tree, it incrementally builds only the relevant portion based on the heuristics. This approach enables efficient pruning and supports fully parallel exploration of independent branches. The resulting partial ordering can then be passed to the symbolic timing tool to compute the corresponding delay expression.

\section{Conclusion}
We presented a symbolic timing analysis tool for digital circuits based on recently developed analytic delay formulas for NOR gates. Our approach composes closed-form $\delta(T,\Delta)$ expressions throughout a circuit, given some specific transition ordering, and computes the output transition times symbolically. Applied to the NOR-based \texttt{c17} slack benchmark, our tool successfully propagated symbolic delay expressions across multiple gate layers and captured both drafting and MIS effects without requiring analog simulation. The prototype demonstrates the feasibility of symbolic verification at the circuit level, opening the door to further analysis of feedback systems, parameter sweeps, and automated path validation.

\vspace{12pt}

\end{document}